# Large lattice distortions associated with the magnetic transition in $La_{0.7}Sr_{0.3}MnO_3$

F. Weber[1,i], D. Argyriou[2], O. Prokhnenko[2], and D. Reznik[1,3,ii,iii]

[1] *Institute of Solid State Physics, Karlsruhe Institute of Technology, 76021 Karlsruhe, Germany*
[2] *Helmholtz-Zentrum Berlin für Materialien und Energie, Hahn-Meitner-Platz 1, 14109 Berlin, Germany*
[3] *Laboratoire Léon Brillouin (CEA-CNRS),CEA-Saclay, F-91911 Gif-sur-Yvette, France*

**Colossal magnetoresistance (CMR) is associated with the phase transition from a metallic ferromagnetic to insulating paramagnetic phase, which can be controlled by an applied magnetic field. The insulating phase occurs due to trapping of the charge carriers by polaronic lattice distortions, which raise the resistivity. Theories based on local physics predict that the magnitude of the resistivity jump at $T_c$ is determined by how much, on average, the amplitude of these distortions increases at the phase transition. Using neutron scattering, we measured the average distortion amplitude in $La_{0.7}Sr_{0.3}MnO_3$. Surprisingly, its increase from below to above $T_c$ is just as large as in other manganites, which have a much larger resistivity jump. This result suggests that the strength of CMR is determined not by the size of distortions, but by their cooperative nature specific to each compound. Existing theories need to be extended to include correlations between different unit cells to explain and predict the strength of CMR.**

[i] Email: frank.weber@kit.edu
[ii] Current address: Department of Physics, University of Colorado at Boulder, Boulder, Colorado 80309, USA
[iii] Email: dmitry.reznik@colorado.edu

Manganese perovskite oxides have a rich temperature-doping phase diagram spanning a wide range of physical phenomena of both practical and fundamental interest such as metal-insulator transitions, colossal magnetoresistance, multiferroicity, polaron formation etc. This behavior is driven by strong coupling between magnetic, orbital, charge, and atomic lattice degrees of freedom with ordering in one channel having a strong, often transformational effect on the other. Electron-phonon coupling through the Jahn-Teller effect plays a special role by tying electronic orbital states to the atomic lattice whose deformations trap charge carriers and thus raise the electrical resistivity. In half-doped manganites the Jahn-Teller effect leads to CE order characterized by very large resistivity, which then drops by several orders of magnitude when the CE order melts at elevated temperatures.

At lower doping Jahn-Teller distortions become frustrated: A distortion in one unit cell interferes with the distortion in the neighboring unit cells and long range CE order cannot form. This part of the phase diagram is famous for its outstandingly large (colossal) magnetoresistance (CMR)[6],[7]. CMR occurs due to the ferromagnetic transition accompanied by a large resistivity change. The resistivity is metallic-like at temperatures, T, less than the ferromagnetic transition temperature, $T_C$, i.e. it is relatively low and increases with T, whereas at $T > T_C$ it is very large and typically decreases with T. Due to high sensitivity of $T_C$ to the applied magnetic field, resistivity changes dramatically as a function of field in the vicinity of $T_C$. The CMR mechanism itself is straightforward, but what is unusual is the very existence of the "metal-insulator" transition at $T_C$.

It has been proposed that the resistivity increases because of competition between double-exchange and Jahn-Teller interactions [8,9]. The former favors ferromagnetism and an undistorted lattice and wins below $T_C$. The latter favors polaronic lattice distortions and paramagnetism or antiferromagnetism and wins above $T_C$ [9]. Theoretical models make the approximation that only local lattice distortions are important[9], and the resistivity scales with $U^2$, the mean square displacement of oxygen atoms from the average periodic position. A jump in $U^2$ at $T_C$ should reflect the strength of the observed CMR effect. This theory explains the shape and temperature dependence of the resistivity curve. In order to design new CMR materials whose properties are better suited for applications as well as from the fundamental science point of view, theory must also relate microscopic parameters derived from chemistry and structure to the magnitude of the resistivity jump at $T_C$ as well as to the value of $T_C$ itself. In existing theories tuning parameters determine the strength of the electron-phonon coupling, which, in turn, determines the amplitude of the polaronic contribution to $U^2$. The larger the polaronic $U^2$, the deeper the trapping potential. Increased trapping potential reduces $T_C$ and increases the strength of CMR. This mechanism is supposed to explain the decrease in ferromagnetic transition temperatures and increased magnetoresistance going from $La_{0.7}Sr_{0.3}MnO_3$ ($T_C$=355 K) to $La_{0.75}Ca_{0.25}MnO_3$ ($T_C$=240 K) and $La_{1.2}Sr_{1.8}Mn_2O_7$ ($T_C$=120 K). On the other hand, the picture of strong electron-lattice interaction has been questioned for thin films of $La_{1-x}Ca_xMnO_3$, $x \geq 0.5$ [10].

We used neutron scattering to measure temperature-dependence of $U^2$ in $La_{0.7}Sr_{0.3}MnO_3$ in order to investigate the relationship between $U^2$ and the electrical resistivity known from the literature. $La_{0.7}Sr_{0.3}MnO_3$ has a relatively small resistivity increase at $T_C$, remains metallic even above $T_C$, and is viewed as a canonical double-exchange compound. We observed a strong increase of $U^2$ at $T_C$, which is evidence for a strong electron-lattice interaction in $La_{0.7}Sr_{0.3}MnO_3$. In fact, the jump in $U^2$ at $T_C$ is similar to that in compounds with much larger magnetoresistance, i.e. variations in the strength of the CMR effect between different compounds are not accompanied by a corresponding variation in the jump of $U^2$ at $T_C$.

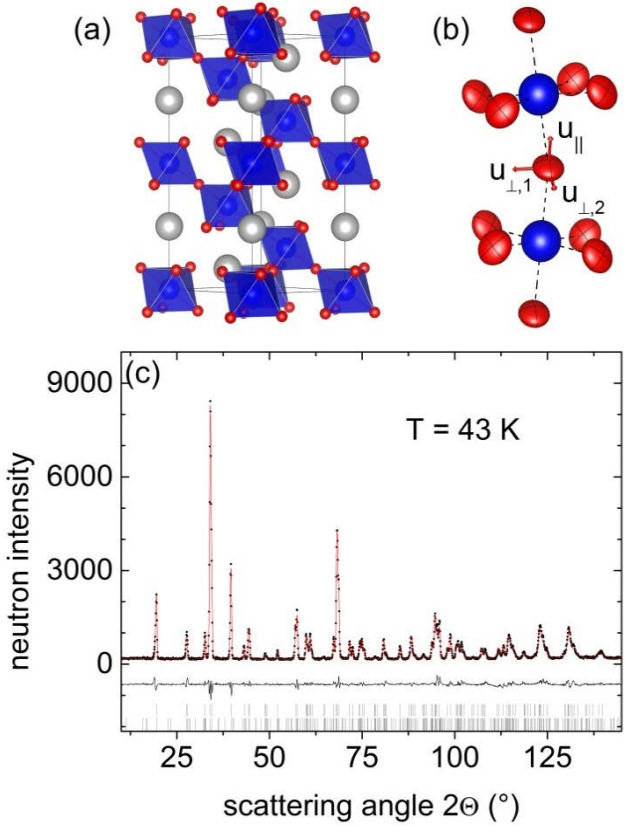

Figure 1. *(a)* Crystal structure of $La_{0.7}Sr_{0.3}MnO_3$ as refined at T = 43 K from neutron powder diffraction. Shown are the $MnO_6$ polyhedra and La/Sr atomic positions. Lines denote the hexagonal unit cell (a = b = 5.479Å, c = 13.276 Å). *(b)* Two consecutive $MnO_6$ octahedra, where oxygen atoms are replaced by the thermal ellipsoid as determined from neutron diffraction at T = 43 K. The principal axes of the ellipsoid are indicated by the black lines on the surface and directions corresponding to the labels of Fig. 2 are shown as red errors. For the sake of clarity, we plot enlarged ellipses (factor of 2). *(c)* Rietveld refinement patterns for $La_{0.7}Sr_{0.3}MnO_3$ at T = 43 K. Dots are observed neutron diffraction intensities, and the solid line represents the calculated fit. The curve below is the difference between the observed and calculated intensities. The upper and lower sets of vertical bars indicate the positions of Bragg reflections for the magnetic and atomic structure, respectively.

Quantitatively, $U^2$ enters the Debye-Waller (DW) factor $e^{-2W}$, $2W = -U^2$ [11], that can be obtained directly from powder diffraction spectra as well as phonon intensity measurements performed on single crystals. We measured powder diffraction spectra of $La_{0.7}Sr_{0.3}MnO_3$ on the high resolution powder diffractometer E9 "Firepod" at Helmholtz center Berlin (HZB) between T = 43 K (Fig. 1) and 530 K using a Ge(711) monochromator and neutrons of wave length 1.30782 Å. Phonons were measured in $La_{1.2}Sr_{1.8}Mn_2O_7$, $La_{0.8}Sr_{0.2}MnO_3$, and $La_{0.7}Sr_{0.3}MnO_3$ on the 1T double focusing neutron triple-axis spectrometer located at the ORPHEE research reactor at the Laboratoire Léon Brillouin (LLB), CE Saclay, using a Cu220 monochromator and PG002 analyzer with final energy fixed at 13.4, 14.7 or 30.5 meV. The sample was mounted in a closed cycle refrigerator capable of reaching high temperatures.

In our analysis of the diffraction data we considered anisotropic $U^2$ for the oxygen atom, i.e. displacements primarily parallel to the Mn-O bond $U_{\parallel}^2$ were treated separately from the two directions perpendicular to the Mn-O bond $U_{\perp,1}^2$ and $U_{\perp,2}^2$ (see Fig. 1b). In this case, the mean square atomic displacements are

$$U^2 = \tfrac{1}{3}\left(U_{\parallel}^2 + U_{\perp,1}^2 + U_{\perp,2}^2\right). \quad (1)$$

Fig. 1a shows our analysis for the anisotropic components of the mean square displacement $U^2$ of oxygen ions in $La_{0.7}Sr_{0.3}MnO_3$ in eq. (1). Indexing of the components is relative to the Mn-O bond, i.e. $U_{\parallel}^2$ is along the Mn-O bond, $U_{\perp,1}^2$ is perpendicular to the Mn-O bond with a c-axis component and $U_{\perp,2}^2$ is perpendicular to the Mn-O bond within the ab plane of the hexagonal unit cell. Figure 1c shows powder diffraction pattern measured on our sample together with Rietveld refinement from which the components of the lattice distortion were extracted using standard procedure.

$U^2$ may be finite due to structural disorder, polaronic lattice distortions connected to CMR and/or thermal motion of the atoms. In order to look for a connection between lattice distortions and CMR, it is necessary to separate polaronic distortions from the others. Distortions due to structural disorder are temperature-independent, whereas the other two components vanish at zero temperature and increase on heating. We calculated the temperature-dependence of the thermal motion contribution to $U^2$ using a shell model based on extensive phonon measurements in $La_{0.7}Sr_{0.3}MnO_3$ performed by Reichardt and Braden [12]. Subtracting the results of this calculation plus a constant term equal to the zero temperature intercept from the data allowed us to obtain the temperature dependence of polaronic distortions associated with CMR. The shell model describes $U_{\parallel}^2$ and $U_{\perp,1}^2$ components of $U^2$ well below $T_C$, however it seems to underestimate the increase in $U_{\perp,2}^2$ for an unknown reason (Fig. 2a).

Fig. 2a shows that all three components of $U^2$ jump at the FM phase transition and then continue to increase as before. $U_{\parallel}^2$ has the sharpest increase at $T_C$, as expected from the formation of Jahn-Teller distortions. We can extract the contribution of polaronic lattice distortions to $U^2$ by subtracting the shell model curve, $U_{model}^2$, from the experimental values, $U_{exp}^2$.

The difference $U_{exp}^2 - U_{model}^2$ versus $T/T_C$ are plotted in Fig. 2b for a number of compounds. As no models were available for the thermal motion of the Ca-doped compounds, we used the respective calculations for $La_{0.7}Sr_{0.3}MnO_3$. The fact that the high temperature values of $U^2_{exp} - U^2_{model}$ are nearly constant supports this procedure. The increase in $U^2_{exp} - U^2_{model}$ from low temperatures to temperatures above $T_C$ is similar in all four compounds. Fig. 2c illustrates the absence of a correlation between the magnitude of jump in the lattice distortion at $T_C$ and the strength of the CMR effect. In the case of a linear correlation, the points would fall on the straight line, but the value for $La_{0.7}Sr_{0.3}MnO_3$ is clearly not following such a trend.

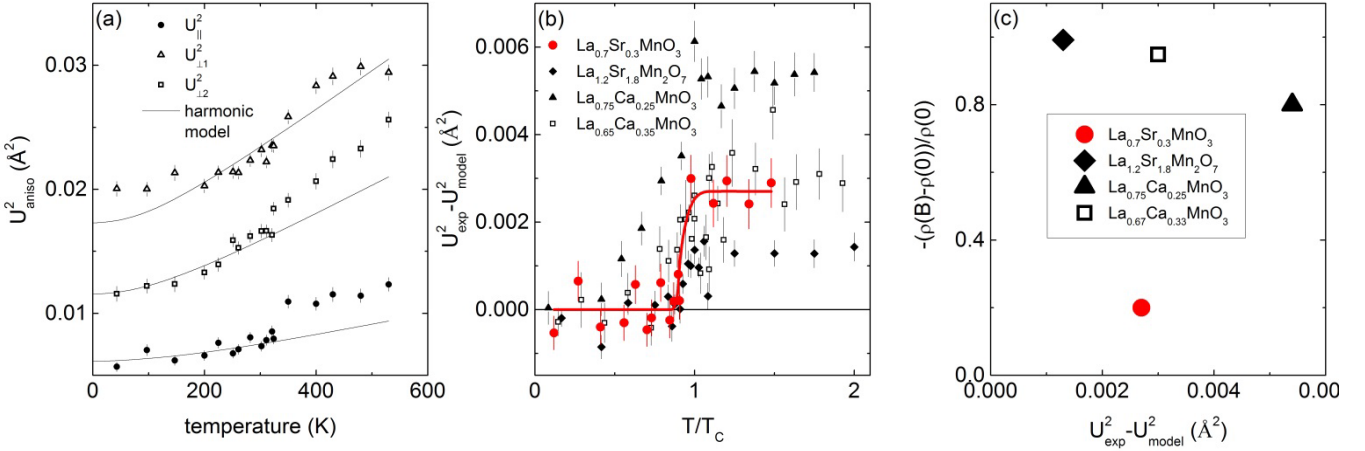

Figure 2. *(a)* Anisotropic oxygen mean square displacements as function of temperature in $La_{0.7}Sr_{0.3}MnO_3$. Dots/lines represent the experimental/calculated values pointing approximately (see Fig. 1b) along the Mn-O bond $U^2_\parallel$ (circles), perpendicular to the Mn-O bond with a c-axis component $U^2_{\perp,1}$ (triangles) and perpendicular to the bond in the hexagonal a-b plane $U^2_{\perp,2}$ (squares). Offsets of 0.003 Å$^2$ and 0.006 Å$^2$ are included in the experimental data for $U^2_{\perp,2}$ and $U^2_{\perp,1}$, respectively. Calculated values were shifted in order to fit the observed low temperature values, where possible. *(b)* Comparison of $U^2_{exp}$-$U^2_{model}$ (see text) for $La_{0.7}Sr_{0.3}MnO_3$ (circles), $La_{1.2}Sr_{1.8}Mn_2O_7$ (diamonds), $La_{0.75}Ca_{0.25}MnO_3$ (triangles) and $La_{0.65}Ca_{0.35}MnO_3$ (squares). $U^2_{exp}$ are the measured values of $U^2$: $U^2_{exp} = U^2_\parallel$ ($La_{0.7}Sr_{0.3}MnO_3$, $La_{0.75}Ca_{0.25}MnO_3$) and $U^2_{exp} = U^2_{iso}$ ($La_{1.2}Sr_{1.8}Mn_2O_7$, $La_{0.65}Ca_{0.35}MnO_3$). Data of the last three compounds are taken from Refs.[1]. For the Ca doped manganites no shell model was available so corresponding calculations for the Sr doped compounds were used for the difference. Line is guide to the eye. *(c)* Magnetoresistance $-(\rho(B) - \rho(0))/\rho(0)$ near $T_c$ of the respective compound in an applied field B = 4 T – 5 T (taken from Refs. [2-4]) versus the average deviations of oxygen mean square displacements (see panel (b)).

Another way of studying oxygen distortions is to measure coherent oxygen vibrations (phonons) by inelastic neutron scattering. The main effect of increased disorder on heating is to reduce intensities of coherent phonons via the Debye-Waller (DW) factor [11]. Understanding this intensity reduction for all phonons is not straightforward because it probably involves both the loss of coherence and eigenvector changes. However, zone center Mn-O bond-stretching phonons are not affected by the latter, because their eigenvectors are constrained by crystal symmetry. After correcting for trivial temperature dependence due to the Bose factor, any coherent intensity reduction at the zone center as a function of temperature results from the increase in $U^2$. Such intensity reduction in a bond-stretching branch was first reported and explained within a similar framework for $La_{0.7}Ca_{0.3}MnO_3$ by Zhang et al.[13].

Fig. 3a-c compares changes in bond-stretching phonon intensity across $T_c$ in $La_{1.2}Sr_{1.8}Mn_2O_7$, $La_{0.7}Sr_{0.3}MnO_3$ and $La_{0.8}Sr_{0.2}MnO_3$ measured in identical experimental conditions. The changes in the phonon intensity and, therefore, in the lattice coherence, are very similar, despite the much smaller resistivity jump at $T_c$ in $La_{0.7}Sr_{0.3}MnO_3$.

Fig. 3d summarizes the results on phonon intensities in a plot of the phonon intensity normalized to the respective low temperature values versus $T/T_C$. We observe a clear jump at $T = T_C$. On the other hand, data below and above $T_C$ for all compounds show a very similar behavior. Hence, the increase in the oxygen Debye-Waller factor in these compounds is very similar as well.

Qualitatively, one can understand the observed behavior based on the orbital structure of Mn sites in perovskite manganites. In $(La_{1-x},(Sr/Ca)_x)_{n+1}Mn_nO_{3n+1}$, there are 4-x electrons per manganese (where x is doping) with the orbital split into a lower lying $t_{2g}$ triplet and a higher energy $e_g$ doublet. Strong on-site repulsion ensures no double occupation of orbitals occurs and Hund's rule coupling aligns all spins on the same Mn site ferromagnetically. Thus the $t_{2g}$ triplet acts as a core orbital occupied by three electrically inert electrons whose spins are aligned into a spin 3/2 total moment. The $e_g$ orbitals may be either singly occupied or empty with a significant intersite overlap allowing hopping through the lattice. They form a conduction band at $1 > x > 0$. The hopping matrix element is determined by the Mn-O-Mn bond angle. It also depends on the alignment of the core spins: The constraint that the spins of $e_g$ electrons must be parallel to the core spin makes the hopping easier if all core spins are aligned ferromagnetically. Thus, ferromagnetic alignment of the $t_{2g}$ spins increases the bandwidth of the conduction electrons and is favored at low temperatures. This effect is called the double exchange interaction [14]. As the bandwidth is reduced above $T_C$, polaronic lattice distortions due to the Jahn-Teller effect appear and trap conduction electrons thus increasing the resistivity.

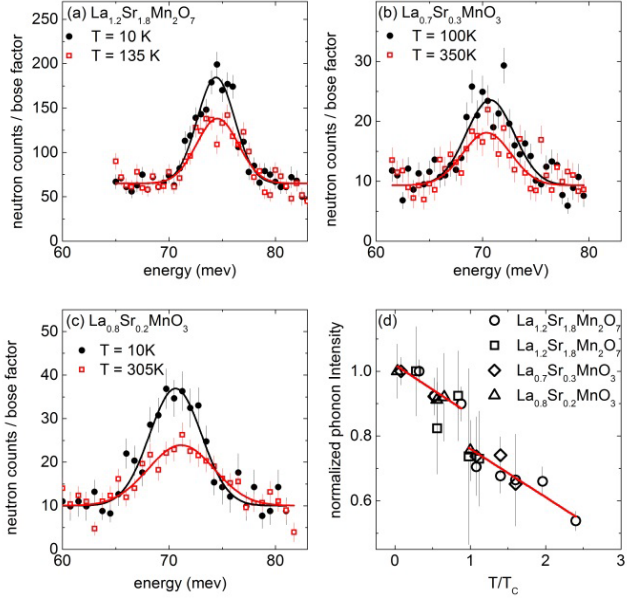

Figure 3. Zone center Mn-O bond stretching modes in (a) $La_{1.2}Sr_{1.8}Mn_2O_7$, (b) $La_{0.7}Sr_{0.3}MnO_3$ and (c) $La_{0.8}Sr_{0.2}MnO_3$ well below (circles) and close (squares) to the respective Curie temperatures. Raw data were corrected for the Bose factor. Constant offsets were subtracted from the high temperature data for a better comparison of the temperature dependent peak intensities. Lines are Gaussian fits of the neutron data. (d) Intensities of zone center Mn-O bond stretching phonons normalized to their low temperature values in $La_{0.8}Sr_{0.2}MnO_3$ (triangles), $La_{0.7}Sr_{0.3}MnO_3$ (routes) and $La_{1.2}Sr_{1.8}Mn_2O_7$ (squares) across the respective Curie temperatures $T_C$ plotted versus $T/T_C$. In $La_{1.2}Sr_{1.8}Mn_2O_7$ we also plot the evolution for the Mn-O bond-bending mode observed at $E = 43$ meV [5] (circles). Lines are linear fits for all points $T < T_C$ and $T \geq T_C$, respectively.

How to explain the differences in the resistivity jump at $T_C$ between different compounds? One possibility is that the magnitude of lattice distortions is larger in compounds with the bigger resistivity jump as proposed in Refs. [9,15]. In this approach, the differences in magnitudes of magnetoresistance between different compounds can be explained by competition between the intersite hopping matrix element, $t$, of $e_g$ electrons and the strength of electron-phonon Jahn-Teller coupling, $\gamma$, which is responsible for different values of $U^2$ in different compounds. Models of this type make an approximation that only necessitates considering the jump in the average displacement $U^2$ of O ions from their crystallographic positions at $T_C$ to capture the relevant physics. In this model the magnitude of the resistivity jump at $T_C$ in different compounds should scale with the jump in $U^2$.

However, this prediction clearly contradicts our results, which show that $U^2$ increases by similar amounts at $T_C$ in manganites with very different resistivity jumps:

$La_{0.7}Sr_{0.3}MnO_3$ is considered to be a "canonical" ferromagnetic manganite with the $T_C$ of 358K and its resistivity increases smoothly through $T_C$ reaching 7.5 x10$^{-3}$ $\Omega$cm [2] at 400K. $La_{0.8}Sr_{0.2}MnO_3$ has a lower $T_C$ (305K) and the resistivity above $T_C$ of 0.01 $\Omega$cm[2], $La_{0.65}Ca_{0.35}MnO_3$ has a $T_C$ of 275K and the resistivity above $T_C$ of 0.04 $\Omega$cm[4], $La_{0.75}Ca_{0.25}MnO_3$ has a $T_C$ of 240 K and a resistivity above $T_C$ of 0.1 $\Omega$cm, whereas in $La_{1.2}Sr_{1.8}Mn_2O_7$ the $T_C$ is 120K and the resistivity above $T_C$ is 0.1-0.2 $\Omega$cm[16]. Fig. 2c illustrates that also the magnetoresistance is very different in these compounds.

Further studies have shown that there is relatively little variation in other anomalous properties between manganites with the ferromagnetic ground state. For example magnon dispersion anomalies are similar in Sr and Ca-doped manganites at the same doping concentration, although the resistivity jump at $T_C$ is much lower and the $T_C$ is much higher in the former [17]. Also, evidence for dynamic polarons was reported in $La_{0.7}Sr_{0.3}MnO_3$ [18].

Our results suggest that instead of the magnitude of the polaronic distortions, the magnitude of the resistivity jump at $T_C$ must be controlled by the dynamics of these distortions: The larger their diffusion rate, the smaller the resistivity. Dynamics of the distortions must be controlled in turn by how distortions in neighboring unit cells are correlated. This suggestion is supported by the fact that CMR effect is the strongest where static or quasistatic short range CE order forms in the paramagnetic phase. This static order must be much more effective in trapping the charge carriers in CMR compounds than equally large dynamic polaronic distortions in $La_{0.7}Sr_{0.3}MnO_3$ where no static order appears.

In order to extend theoretical models of CMR to include dynamics of the distortions, it is necessary to go beyond local physics and consider correlations between distortions in different unit cells. Another potentially promising approach is to try to better understand electron-phonon coupling from the point of view of phonons, i.e. which phonon modes are affected by the CMR effect and which are not. In this approach the correlated motion of atoms is built in from the start.

Our results, combined with earlier studies, show that electron-phonon coupling is strong in all manganites with the ferromagnetic ground state and its strength by itself determines neither the $T_C$ nor the magnitude of the resistivity jump at $T_C$. Thus it is necessary to reexamine the current theoretical underpinning of the CMR phenomenon, which predicts a direct relationship between these quantities. We believe that the key to understanding CMR is in relatively subtle effects such as intersite correlations and/or dynamics of lattice distortions. The rigidity of the distortions must be the dominant force behind the resistivity increase above $T_C$. A successful theory must be able to derive this mechanism from materials' structure and chemistry.

1. D. N. Argyriou, H. N. Bordallo, J. F. Mitchell, J. D. Jorgensen, and G. F. Strouse, Phys. Rev. B **60**, 6200 (1999); P. G. Radaelli, G. Iannone, M. Marezio, H. Y. Hwang, S.-W. Cheong, J. D. Jorgensen, and D. N. Argyriou, Phys. Rev. B **56**, 8265 (1997); P. Dai, J. Zhang, H. A. Mook, S.-H. Liou, P. A. Dowben, and E. W. Plummer, Phys. Rev. B **54**, R3694 (1996).
2. A. Urushibara, Y. Moritomo, T. Arima, A. Asamitsu, G. Kido, and Y. Tokura, Phys. Rev. B **51**, 14103 (1995).
3. P. Schiffer, A. P. Ramirez, W. Bao, and S.-W. Cheong, Phys. Rev. Lett. **75**, 3336 (1995).
4. Y. Tokura, *Collossal magnetoresistive oxides*. (2000).
5. F. Weber, N. Aliouane, H. Zheng, J. F. Mitchell, D. N. Argyriou, and D. Reznik, Nature materials **8** (10), 798 (2009).
6. R. v. Helmholt, J. Wecker, B. Holzapfel, L. Schultz, and K. Samwer, Phys. Rev. Lett. **71**, 2331 (1993).
7. S. Jin, T. H. Tiefel, M. McCormack, R. A. Fastnacht, R. Ramesh, and L. H. Chen, Science **264**, 413 (1994).
8. A. J. Millis, P. B. Littlewood, and B. I. Shraiman, Phys. Rev. Lett. **74**, 5144 (1995).
9. A. J. Millis, R. Mueller, and B. I. Shraiman, Phys. Rev. Lett. **77**, 175 (1996).
10. J. Loudon, S. Cox, A. Williams, J. Attfield, P. Littlewood, P. Midgley, and N. Mathur, Physical Review Letters **94** (9), 097202 (2005); S. Cox, J. Singleton, R. D. Mcdonald, A. Migliori, and P. B. Littlewood, Nature materials **7**, 25 (2007).
11. G. Squires, *Introduction to the Theory of Thermal Neutron Scattering*. (1978).
12. W. Reichardt and M. Braden, Physica B **263-264**, 416 (1999).
13. J. Zhang, P. Dai, J. A. Fernandez-Baca, E. W. Plummer, Y. Tomioka, and Y. Tokura, Phys. Rev. Lett. **86**, 3823 (2001).
14. C. Zener, Phys. Rev. **81**, 440 (1951); C. Zener, Phys. Rev. **82**, 403 (1951); J. B. Goodenough, Phys. Rev. **100**, 564 (1955); P.-G. d. Gennes, Phys. Rev. **118**, 141 (1960).
15. A. C. M. Green, Phys. Rev. B **63** (20), 205110 (2001).
16. X. J. Chen, C. L. Zhang, C. C. Almasan, J. S. Gardner, and J. L. Sarrao, Phys. Rev. B **67**, 094426 (2003).
17. P. Dai, H. Y. Hwang, J. Zhang, J. A. Fernandez-Baca, S.-W. Cheong, C. Kloc, Y. Tomioka, and Y. Tokura, Phys. Rev. B **61**, 9553 (2000).
18. Y. Chen, B. G. Ueland, J. W. Lynn, G. L. Bychkov, S. N. Barilo, and Y. M. Mukovskii, Physical Review B **78** (21), 212301 (2008).